\documentclass{ws-procs10x7}
\include{epsfig}
\newcommand{\oaa}    {\ensuremath{\mathcal{O}(\alpha_{\mathrm{s}}^2)}}
\newcommand{\oaaa}    {\ensuremath{\mathcal{O}(\alpha_{\mathrm{s}}^3)}}
\newcommand{\dd}    {\ensuremath{\mathrm{d}}}
\newcommand{\as}     {\ensuremath{\alpha_{\mathrm{s}}}}
\begin{document}

\title{QCD Studies and $\bm\alpha_{\rm s}$ Measurements at LEP}

\author{Stefan S\"oldner-Rembold}

\address{University of Manchester, Oxford Road, M13 9PL, Manchester,
UK\\E-mail: soldner@fnal.gov}

\twocolumn[\maketitle\abstract{
The LEP experiments have measured event shapes using data taken at 
e$^+$e$^-$ center-of-mass energies $\sqrt{s}$ ranging from 91 GeV to 209 GeV.
Using the final LEP event shape measurements, a combined value of 
the strong coupling constant $\alpha_{\rm s}(M_{\rm Z})$
has been extracted.
Events with photon radiation have been used to extend the measurements
to lower center-of-mass energies $\sqrt{s'}$ to study the running of
$\alpha_{\rm s}$. Alternative \as\ measurements using four-jet rates
have also been performed. 
}]

\section{Data Sample}
Between 1989 and 2000 the four LEP experiments have taken data at 
e$^+$e$^-$ center-of-mass energies $\sqrt{s}$ ranging from 91 GeV to 209 GeV.
Each experiment has collected several million
multi-hadronic events at $\sqrt{s}=M_{\rm Z}$.
At higher energies about $10^{3}$ events
have been recorded per energy point. After a basic event
selection, no significant background remains
at $\sqrt{s}=M_{\rm Z}$. In the range $\sqrt{s}>2M_{\rm W}$ the
background rate is about
$10-15\%$ due to four-fermion production in
$\mbox{e}^+\mbox{e}^-\to\mbox{W}^+\mbox{W}^-$ events.
For an accurate determination of $\sqrt{s}$ it is also important
to identify and reconstruct events with initial 
and final state photon radiation. 
\vspace{-2mm}
\section{Event Shape Observables}
The properties of hadronic events may be described by a set of
event shape observables. These may be used to characterize the
distribution of particles in an event as ``pencil-like'',
planar, spherical, etc.
They can be computed either using the measured charged particles
and calorimeter clusters, or using the hadrons or partons in
simulated events.
In order that predictions of perturbative QCD
be reliable, it is necessary that the value of the observable be 
infra-red stable (i.e.\ unaltered under the emission of soft gluons)
and collinear stable (i.e.\ unaltered under collinear parton branchings).
The following event shapes are considered here:
\begin{description}
\item[Thrust $\bm T$:]
  defined by the expression
$$
  T= \max_{\vec{n}}\left(\frac{\sum_i|p_i\cdot\vec{n}|}
                    {\sum_i|p_i|}\right),
$$
  where $p_i$ is the three-momentum of particle $i$.
  The thrust axis $\vec{n}_T$ is the direction $\vec{n}$ which
  maximizes the expression in parentheses.  A plane through the origin
  and perpendicular to $\vec{n}_T$ divides the event into two
  hemispheres.
\item[$\bm C$-parameter:]
  The linearized momentum tensor is given by
  $$
    \Theta^{\alpha\beta}= \frac{\sum_i(p_i^{\alpha}p_i^{\beta})/|p_i|}
                               {\sum_i|p_i|}
                           \;\;(\alpha,\beta= 1,2,3),
  \nonumber  
$$
  where the sum runs over particles, $i$, and $\alpha,\beta$ denote 
  the Cartesian coordinates of the momentum vector. 
  The three eigenvalues $\lambda_j$ of this tensor define $C$
  through
 $$
    C= 3(\lambda_1\lambda_2+\lambda_2\lambda_3+\lambda_3\lambda_1).
  \nonumber
  $$
\begin{figure*}[htbp]
\epsfxsize30pc
\mbox{\epsfxsize=0.68\columnwidth
 \epsffile{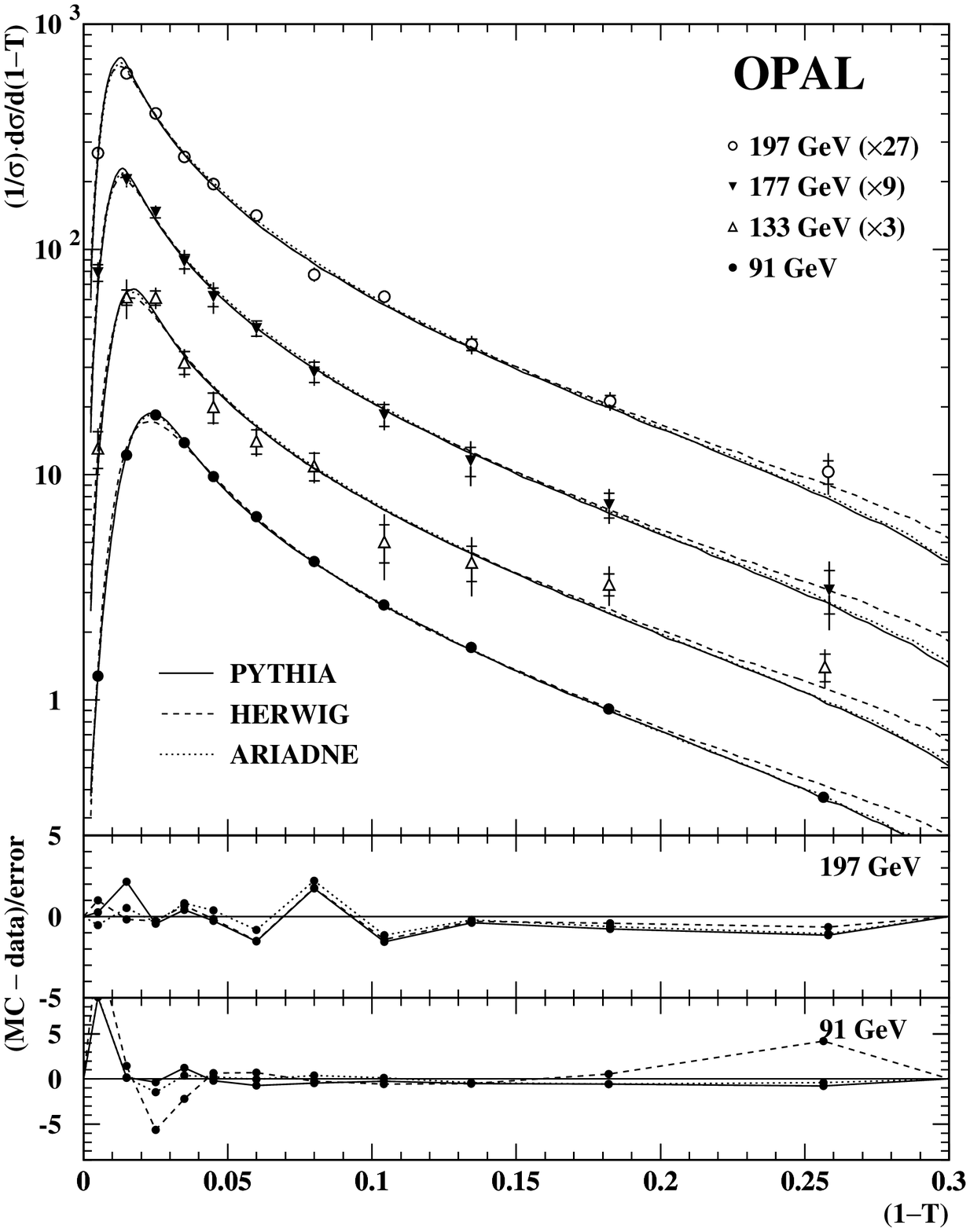}
 \epsfxsize=0.68\columnwidth
 \epsffile{mhigh207.epsi}
 \epsfxsize=0.68\columnwidth
 \epsffile{s03bt.epsi}}
\caption{
(left) $1-T$ measured by OPAL in the range $\sqrt{s}=91-197$~GeV.
(center) $M_H$ measured by DELPHI at $\sqrt{s}=207$ GeV. 
The subtracted four-fermion background is shown separately. 
The OPAL and DELPHI data are compared to PYTHIA (JETSET), HERWIG
and ARIADNE MC simulations.
(right) $B_T$ measured by ALEPH in the range $\sqrt{s}=91-206$~GeV. 
The continuous line shows the region where the
${\cal O}(\as^2)$+NLLA fit has been performed.
\label{fig-lep}}
\end{figure*}
\item[Heavy Jet Mass $\bm M_H$:] The hemisphere 
  invariant masses are calculated using the particles
  in the two hemispheres. $M_H$ is
  the heavier of the two masses.
\item[Jet Broadening $\bm B_T$ and $\bm B_W$:] 
  These are defined by computing the quantity
  $$
    B_k= \left(\frac{\sum_{i\in H_k}|p_i\times\vec{n}_T|}
                    {2\sum_i|p_i|}\right)
  \label{equ_btbw}
  $$
  for each of the two hemispheres.
  The two observables are defined by
  $$
    B_T= B_1+B_2\;\;\;\mathrm{and}
\;\;\;B_W= \max(B_1,B_2),
  $$
  where $B_T$ is the total and
$B_W$ is the wide jet broadening.
\item[Transition between 2- and 3-jets ($\bm y_{23}$):]
  The value of the jet resolution parameter, $y_{\rm cut}$, 
  at which the
  event makes a transition between a  2-jet and a 3-jet assignment,
  for the Durham jet finding scheme~\cite{bib-durham}.
\end{description}

The final results on event shapes have recently been published by
all four LEP collaborations~\cite{bib-aleph,bib-delphi,bib-l3,bib-opal}.
Examples are shown in Fig.~\ref{fig-lep}.
The agreement between data and Monte Carlo (MC) simulations
is in general very good which justifies the use of simple bin-by-bin
efficiency corrections.

The QCD calculations for the cumulative cross-section
$R(y) \equiv \int_0^y \frac{1}{\sigma}\frac{\dd\sigma}{\dd y} \dd y$
as a function of the event shape variable $y (y=1-T,C,M_H,B_T,B_W,y_{23})$
can be performed in the next-to-leading-logarithmic approximation (NLLA)
or using a fixed order \oaa\ calculation.
The \oaa\ prediction is expected to be best for the multi-jet
region (high $y$), whereas the NLLA prediction is best for
the two-jet region (low $y$).
The ``$\log(R)$'' matching 
scheme is adopted for combining 
the \oaa\ and NLLA predictions. 
In the $\log(R)$ matching scheme the terms up to \oaa\ 
in the NLLA expression are replaced
by the \oaa\ terms from $\log R_{\oaa}$.

\begin{figure}[htbp]
   \begin{center}
      \mbox{
          \epsfxsize=0.8\columnwidth
           \epsffile{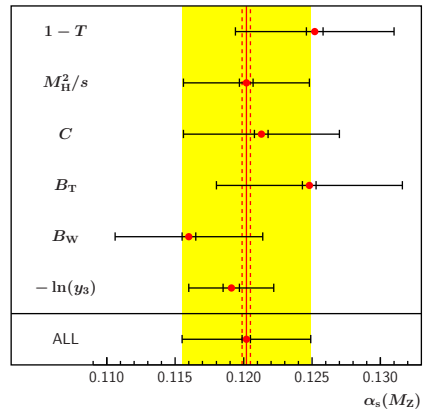}
           }
   \end{center}
\caption{
Combined $\alpha_{\rm s}(M_{\rm Z})$
for the six event shape variables measured
by the four LEP experiments. The inner error bar is the
statistical, 
the outer error bar is the total uncertainty~\protect\cite{bib-ford}.
}
\label{fig-comb1}
\end{figure}

\begin{figure}[htbp]
   \begin{center}
      \mbox{
          \epsfxsize=0.7\columnwidth
           \epsffile{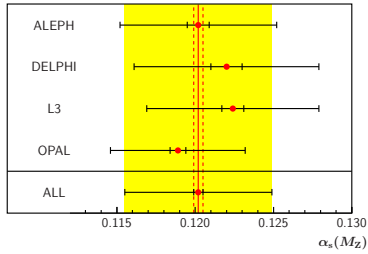}
           }
   \end{center}
\caption{
Combined $\alpha_{\rm s}(M_{\rm Z})$ for the four LEP experiments using the
measurements from the six event shape variables.
The inner error bar is the
statistical, 
the outer error bar is the total uncertainty~\protect\cite{bib-ford}.
}
\label{fig-comb2}
\end{figure}

Each experiment has performed \as\ fits using these predictions
for the six event shape variables at different $\sqrt{s}$.
A preliminary combination of these results, 
taking into account the correlations between the sources of uncertainties,
yields~\cite{bib-ford} 
\begin{eqnarray*}
  \lefteqn{
\alpha_{\rm s}(M_{\rm Z})=}\\
& & 0.1202 \pm 0.0003 (\mbox{stat}) \pm 0.0007 (\mbox{expt})\\
& & \pm 0.0015 (\mbox{hadr}) \pm 0.0044 (\mbox{theo})
\end{eqnarray*}
with $\chi^2/\mbox{d.o.f.}=93/166$.
The combined \as\ measurements for each event shape variable
are shown in Fig.~\ref{fig-comb1} and the combined $\as$ measurements
for each experiment in Fig.~\ref{fig-comb2}. 

\begin{figure}[htbp]
   \begin{center}
      \mbox{
          \epsfxsize=0.7\columnwidth
           \epsffile{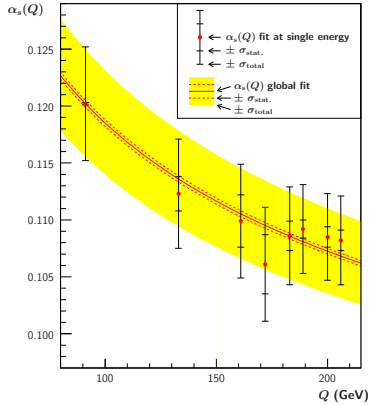}
           }
   \end{center}
\caption{
The running of \as\ as a function of $Q=\sqrt{s}$ using
the combined LEP data.
The inner error bar is the
statistical uncertainty, 
the outer error bar is the total uncertainty~\protect\cite{bib-ford}.
}
\label{fig-comb3}
\end{figure}

\begin{table*}
\caption{Summary of the $\alpha_{\rm s}(M_{\rm Z})$ 
results presented in this talk.}
\label{tab1}
\begin{tabular}{|c|c|c|ccccc|} 
\hline
 & data & theory &  $\alpha_{\rm s}(M_{\rm Z})$ 
& stat & expt & hadr & theo \\
\hline
 LEP & event shapes & \oaa +NLLA  & 
0.1202 & 0.0003 & 0.0007 & 0.0015 & 0.0044 \\
 OPAL & radiative events & \oaa +NLLA  & 
0.1176 & 0.0012 & \multicolumn{3}{c|}{$^{+0.0093}_{-0.0085}$}  \\
 OPAL & $R_4$/Durham & \oaaa +NLLA  & 
0.1208 & 0.0006 & 0.0021 & 0.0019 & 0.0024 \\
 DELPHI & $R_4$/Cambridge & \oaaa +$x_{\mu}^{\rm opt}$  & 
0.1175 & 0.0005 & 0.0010 & 0.0027 & 0.0007 \\
\hline
\end{tabular}
\end{table*}

The experimental uncertainty includes uncertainties from cut
variation, detector corrections and background subtraction.
The hadronization uncertainty is determined by performing the
hadronization correction using different MC generators (HERWIG,
ARIADNE, PYTHIA. The theoretical uncertainty includes
the variation of the renormalization scale, $0.5<x_{\mu}<2$,
the variation of the logarithmic rescaling factor, $2/3<x_{L}<3/2$,
and of kinematic cut-off parameters. In addition, the 
$\log(R)$ matching scheme is replaced by the $R$ matching 
scheme~\cite{bib-jones}. 

The running of \as\ using the combined data is shown in Fig.\ref{fig-comb3}.
Although the statistical precision of the LEP1 data at $\sqrt{s}=M_{\rm Z})$
is much higher, the LEP1 and LEP2 data have about equal weight in
the combined fit because of the smaller hadronization and theory uncertainties
at larger $\sqrt{s}$.

\vspace{-2mm}
\section{$\bm\alpha_{\rm s}$ from Radiative Events}
Radiative hadronic events have been used to extend the range of center-of-mass
energies below $\sqrt{s}=M_{\rm Z}$. Assuming that initial state
or final state photons do not interfere with the 
QCD process~\cite{bib-dasgupta}, a
measurement of \as\ at a reduced center-of-mass energy $\sqrt{s'}$
is possible by identifying isolated high energy photons in the 
detector~\cite{bib-l3rad}.
\begin{figure}[htbp]
   \begin{center}
      \mbox{
          \epsfxsize=0.7\columnwidth
           \epsffile{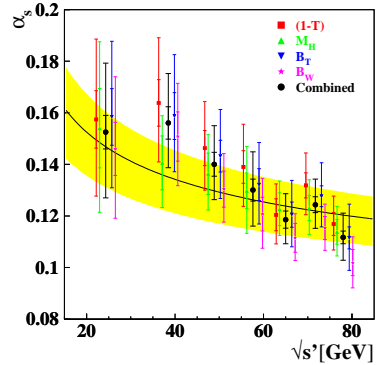}
           }
   \end{center}
\caption{
The running of \as\ as a function of $Q=\sqrt{s'}$ using
radiative hadronic events measured by OPAL~\protect\cite{bib-pn519}.
The result of the \as\ fit is shown as continuous line.
}
\label{fig-opalrad}
\end{figure}

In the analysis performed by OPAL~\cite{bib-pn519} 
photons have been separated from the dominant
$\pi^0$ background by using a likelihood based on cluster shape
fits in the calorimeter. The running of \as\ as function
of $\sqrt{s'}$ is shown in Fig.~\ref{fig-opalrad}.
The result of the \as\ fit is given in Table~\ref{tab1}.

\section{Power Law Corrections}
Non-perturbative effects in event shape observables are
usually suppressed by powers of $1/Q$~\cite{bib-power}.
The corresponding hadronization corrections at LEP energies
are of the order $10\%$. These hadronization corrections
can be determined using MC generators or by analytical 
power law calculations. These calculations introduce one
additional phenomenological parameter $\alpha_0$,
$$
\alpha_0(\mu_I)=\frac{1}{\mu_I}\int_0^{\mu_I}\as(k)\dd k,
$$
which measures the effective strength of the strong coupling
up to an infrared matching scale of $\mu_I\approx 1$~GeV.
The parameter $\alpha_0$ is expected to be universal and must
be determined by experiment.
\begin{figure}[htbp]
   \begin{center}
      \mbox{
          \epsfxsize=0.7\columnwidth
           \epsffile{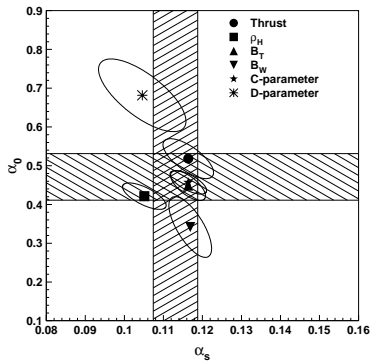}
           }
   \end{center}
\caption{
Contours of confidence levels for simultaneous measurements
of \as\ and $\alpha_0$ measured by L3 using the first moments
of event shape variable. The unweighted averages are shown
as shaded bands. A 4-jet observable (D parameter) is also shown.
}
\label{fig-a0l3}
\end{figure}

\begin{figure}[htbp]
   \begin{center}
      \mbox{
          \epsfxsize=0.7\columnwidth
           \epsffile{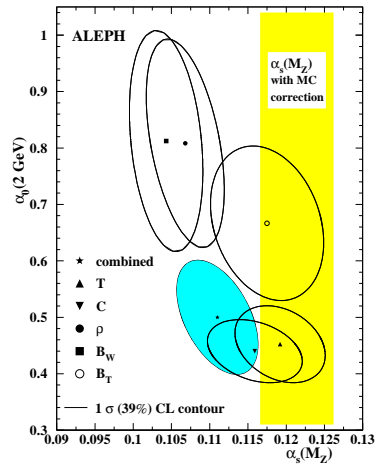}
           }
   \end{center}
\caption{
Contours of confidence levels for simultaneous measurements
of \as\ and $\alpha_0$ compared to the \as\ measurement
using MC corrections (shaded band).
}
\label{fig-a0aleph}
\end{figure}
The result of such a combined fit, performed by L3~\cite{bib-l3}
for the first moments of the event shape
variables, is shown in Fig.~\ref{fig-a0l3}.
The six values of $\alpha_0$ obtained
from the event-shape variables do not agree well. 
A similar result has been obtained by DELPHI~\cite{bib-delphi}.
A large spread of $\alpha_0$ values has also been observed by
ALEPH~\cite{bib-aleph} using event shape distributions. 
The resulting \as\ is significantly
lower than the result obtained using MC hadronization corrections 
(Fig.~\ref{fig-a0aleph}). 

\vspace{-2mm}
\section{$\bm\alpha_{\mathrm s}$ from Four-jet Rates}
\begin{figure}[htbp]
   \begin{center}
      \mbox{
          \epsfxsize=0.7\columnwidth
           \epsffile{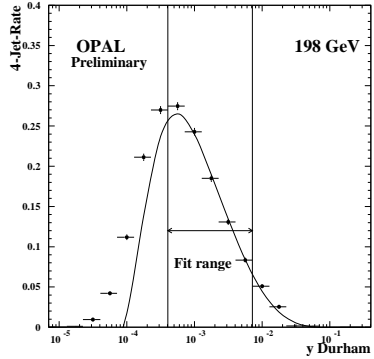}
           }
   \end{center}
\caption{
A fit to the four-jet rate $R_4$ measured at $\sqrt{s}=198$~GeV.
}
\label{fig-r4opal}
\end{figure}

An alternative method to determine \as\ is to measure
four-jet rates, since the jet rates are predicted as
functions of the jet resolution parameter with \as\ as free parameter.
OPAL~\cite{bib-pn527} has measured the four-jet rates using the 
Durham jet finding
algorithm~\cite{bib-durham} as a function of the
jet resolution parameter (Fig.~\ref{fig-r4opal}).
The strong coupling \as\ is fitted using a 
${\cal O}(\as^3)$+NLLA matched calculation of the four-jet rates.

\begin{figure}[htbp]
   \begin{center}
      \mbox{
          \epsfxsize=0.7\columnwidth
           \epsffile{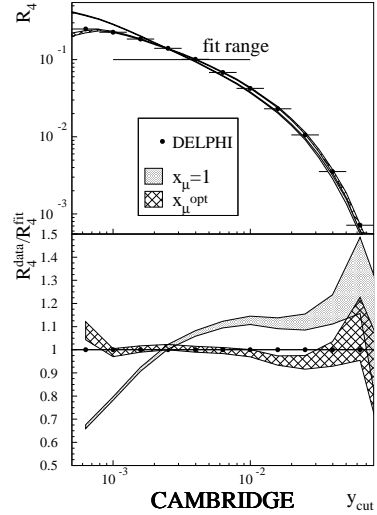}
           }
   \end{center}
\caption{
Fits to the four-jet rates $R_4$ measured at $\sqrt{s}=M_{\rm Z}$
using the Cambridge algorithm.
The grey band shows a fit with the physical scale ($x_{\mu}=1$) and
the cross-hatched band a fit with optimized scales.
}
\label{fig-r4delphi}
\end{figure}

In a similar analysis, DELPHI~\cite{bib-delphi2} has measured the four-jet 
rates
using the Durham and the Cambridge~\cite{bib-cambridge} algorithms.
The data are fitted using a fixed order ${\cal O}(\as^3)$ calculation
with both \as\ and the renormalization scale $x_{\mu}$ as free
parameter. To determine the experimentally optimized scale, a two
parameter fit with \as\ and $x_{\mu}^{\rm opt}$ as free parameters
has also been performed (Fig.~\ref{fig-r4delphi}).
Due to a smaller theoretical uncertainty the \as\ fit using
the Cambridge algorithm is quoted as final result (Table~\ref{tab1}).

\vspace{-2mm}
\section{Summary}
A preliminary \as\ combination using the final 
event shape measurements of all four LEP experiments has been presented.
The universality of power law corrections has been studied.
Radiative events and four-jet events have been used
to study the running of \as.
\vspace{-2mm}
\section*{Acknowledgments}
Special thanks to Matthew Ford for performing
the \as\ combination and 
to Roger Jones, Stefan Kluth, Gavin Salam and Daniel Wicke for their help
in preparing this presentation.

\end{document}